\documentclass[a4paper,10pt]{article}
\usepackage[utf8]{inputenc}
\usepackage{default}
\usepackage{mathtools}
\usepackage{amsmath}

\usepackage{graphicx}
\usepackage{epsfig}
\usepackage{color}
\usepackage{multicol}

\title{Heuristic algorithm for 1D and 2D unfolding.}
\author{Yordan Karadzhov \\ University of Geneva, DPNC \\ yordan.karadzhov@cern.ch}

\begin{document}

\maketitle

\begin{abstract}
A very simple heuristic approach to the unfolding problem will be described. An iterative algorithm 
starts with an empty histogram and every iteration aims to add one entry to this histogram.
The entry to be added is selected according to a criteria which includes a $\chi^2$ test and a regularization.
After a relatively small number of iterations (500 - 1000) the growing reconstructed distribution converges to the
true distribution. 
\end{abstract}

\begin{multicols}{2}

\section{Introduction}
The Linear Inverse Problem or Unfolding is a complex problem common for many experiments. Often the experimentalist
has to reconstruct a true distribution T from a measured distribution M, where the two distributions are connected by

\begin{equation} \label{e1}
 \int R(x,y) T(x) dx = M(y).
\end{equation}
The function $R(x,y)$ can represent the limited resolution and acceptance of the detector, or the presence
of an intermediate process.

In the case of a 1-dimensional (1D) discrete approximation  one can reformulate Eq.\ref{e1} as 
\begin{equation} \label{e2}
 R_{ij}T_j = M_i.
\end{equation}

Here the histograms $T_j \ (j=0,1,2,..,N_t-1)$ and
$M_i \ (i=0,1,2,..,n_m-1)$ are connected by a matrix $R_{ij}$ which  gives the fraction of events
from bin $T_j$ of the true distribution that end up being measured in bin $M_i$ of the measured distribution.
Typically this matrix is determined by a model or by using a Monte Carlo simulation of the direct process.

When solving Eq.(\ref{e2}), $T_j$ and $M_i$ cannot be simply considered as vectors, because the number of entries in a given bin
can only be a non-negative real number.

It is also important to remember that Eq.(\ref{e2}) is an approximation of Eq.(\ref{e1}). The matrix $R_{ij}$ connects
one particular true distribution $T_j$ to one particular measured distribution $M_i$, therefore $R_{ij}$ and $T_j$ are not
independent. A preliminary hypothesis about the true distribution is needed, in order to calculate the elements of $R_{ij}$.
The usage of a wrong hypothesis about T will introduce certain systematic errors in the calculation of this matrix. 

\section{Description of the algorithm}
In principal, if the matrix $R_{ij}$ is already known, one can try to guess the number of entries in every bin of the
true sample, and to use the connection matrix to create a measured sample corresponding to this guess.

\begin{equation}
 R_{ij}T_j^g = M_i^g
\end{equation}

Then the guess $T^g$ can be validated by a comparison between the measured sample $M$ and the
sample $M^g$. A $\chi^2$ test \cite{chi2} can be used for a quantitative estimate of the quality of the guess.
The minimum of $\chi^2$ can also be used as a \textit{selection criteria} for choosing the \textit{best guess}
between multiple candidates. 

A direct brute-force attack\cite{bfa} is not applicable for solving the unfolding problem, because of the unaffordable
number of possible true samples $T^g$, which have to be tested against the measured sample $M$.
Nevertheless, if we limit ourself to the case of guesses $T^g$, containing only one entry, the number of possible
candidates is equal to the number of bins $N_t$, used to depict the true distribution $T$. In this case we can easily select
the \textit{best guess} and there is a good
chance that the entry of this \textit{best guess} will be placed in a bin $T_j^g$ where the Probability Density
Function of the true distribution ($p.d.f._T$) has a relatively big value. Unfortunately, this single entry
cannot be used to derive any useful information about $T$.

\begin{figure}[htb]
 \begin{center}
 \includegraphics[width=.53\textwidth]{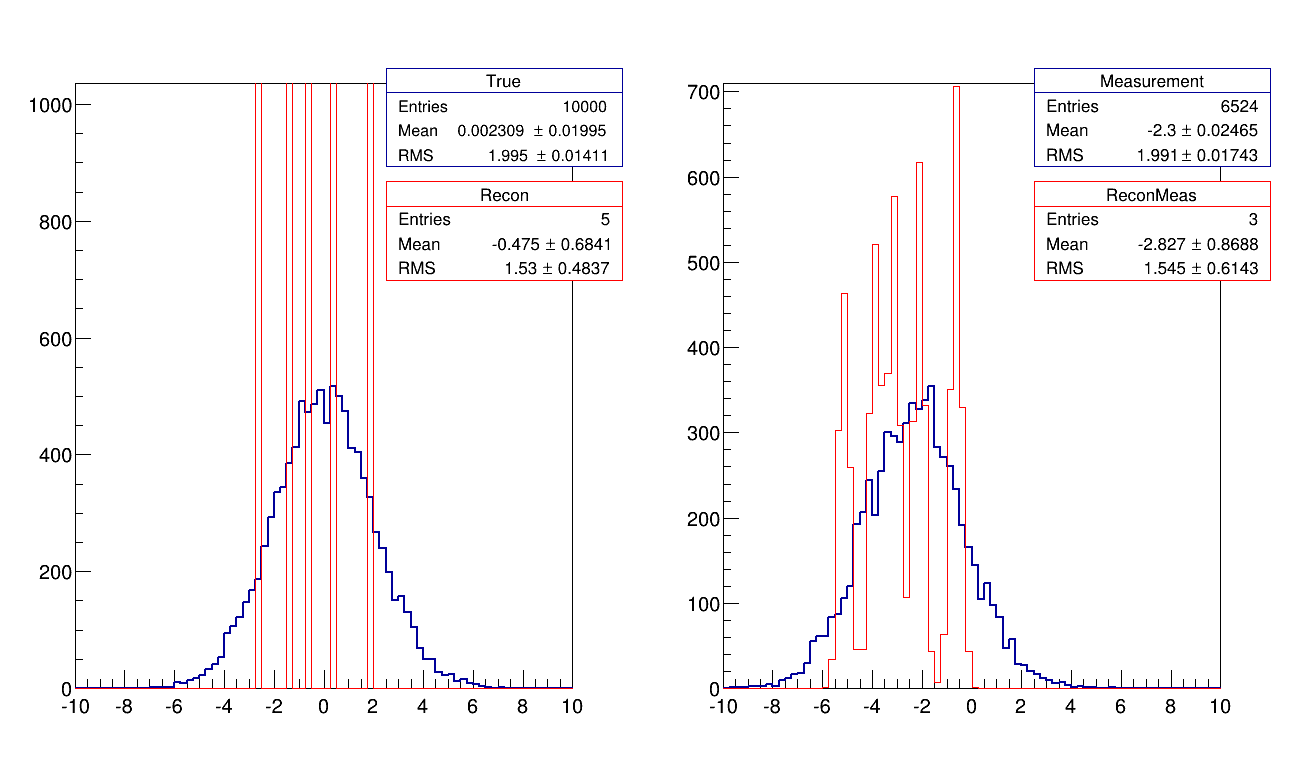}
 \includegraphics[width=.53\textwidth]{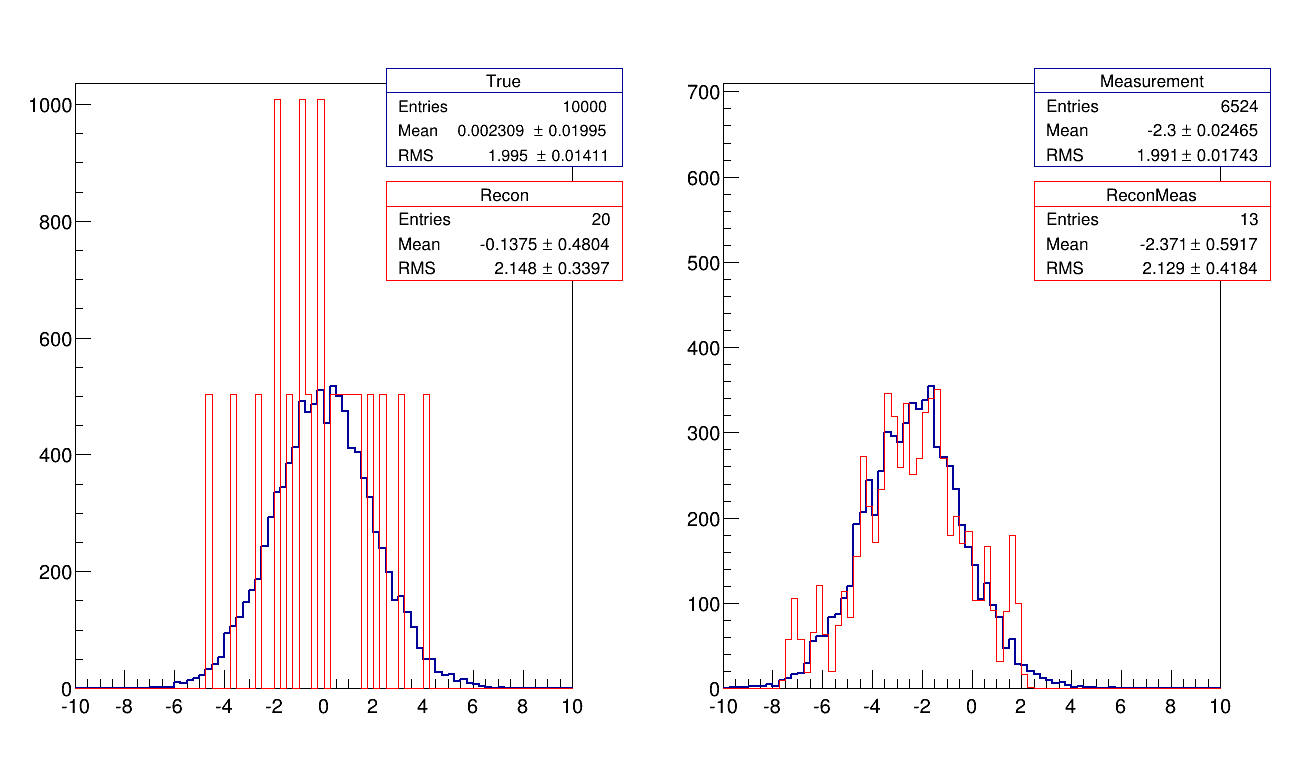}
 \includegraphics[width=.53\textwidth]{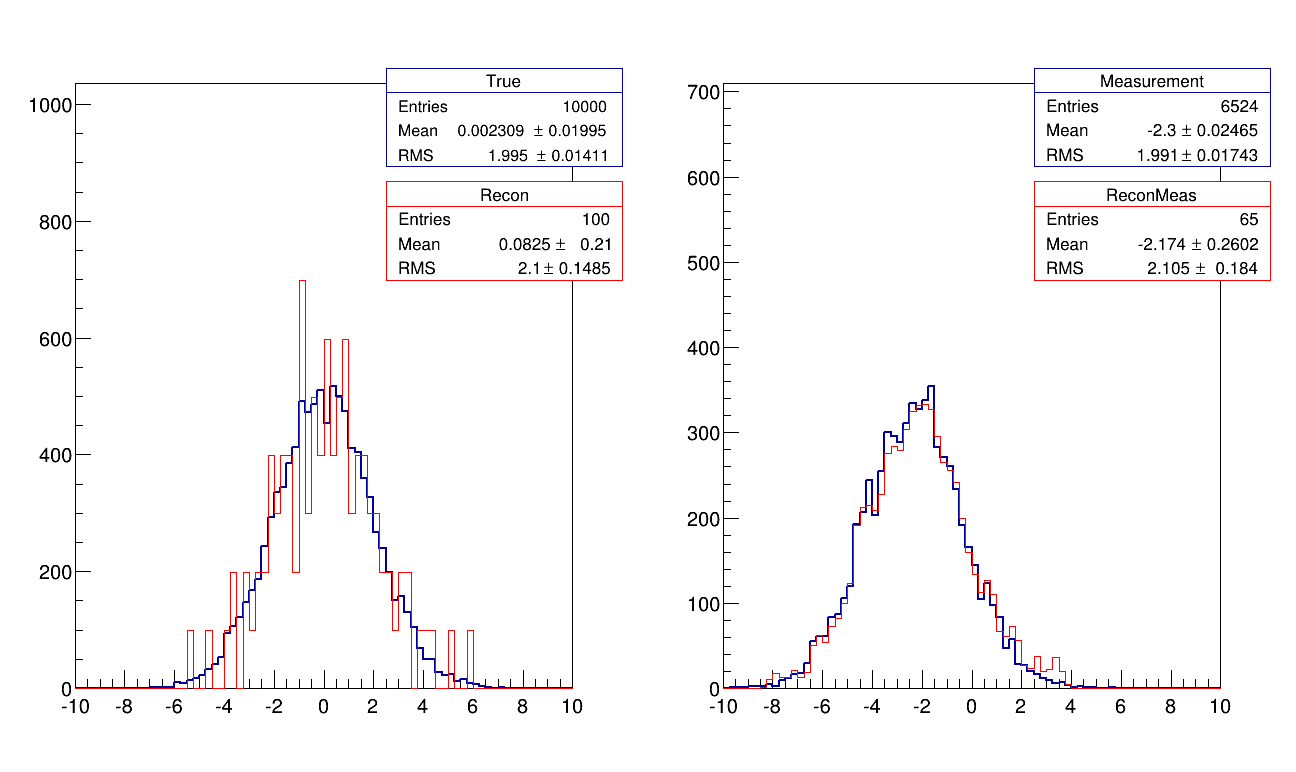}
\end{center}
\caption{The progress of the algorithm after 5, 20 and 100 iterations (top to bottom). 
Left: the true distribution $T$ in blue and the reconstructed distribution $T^g$ (\textit{best guess}) in red.
Right: the measured distribution $M$ in blue and the projection of the reconstructed distribution $M^g$ in red.
The reconstructed distribution and its projection are scaled in order to be compatible with the measured
distribution.}
\label{progres}
\end{figure}

At this stage one can try adding another entry to the \textit{best guess}. This will require a second iteration of the
same procedure, which will include test of $N_t$
new candidates. Again, there is a good chance that the second entry of the \textit{best guess} will
be added to a bin, where $p.d.f._T$ has a relatively big value, but this time the decision will be influenced
also by the choice made during the previous iteration. Every subsequent iteration  of the procedure will add new entry to $T^g$
in a way which gives the best possible match between $M$ and $M^g$. After a sufficient number of iterations the growing
distribution of the \textit{best guess} $T^g$ will start to converge to the true
distribution $T$. This is illustrated in Fig. \ref{progres}.

In this example the connection function $R(x,y)$ corresponds to a gaussian smearing, systematic translation, and
variable inefficiency. The matrix elements of $R_{ij}$ are calculated with Monte Carlo by assuming a flat true distribution.

\section{Regularization of the reconstructed distribution}
\begin{figure}
 \begin{center}
 \includegraphics[width=.53\textwidth]{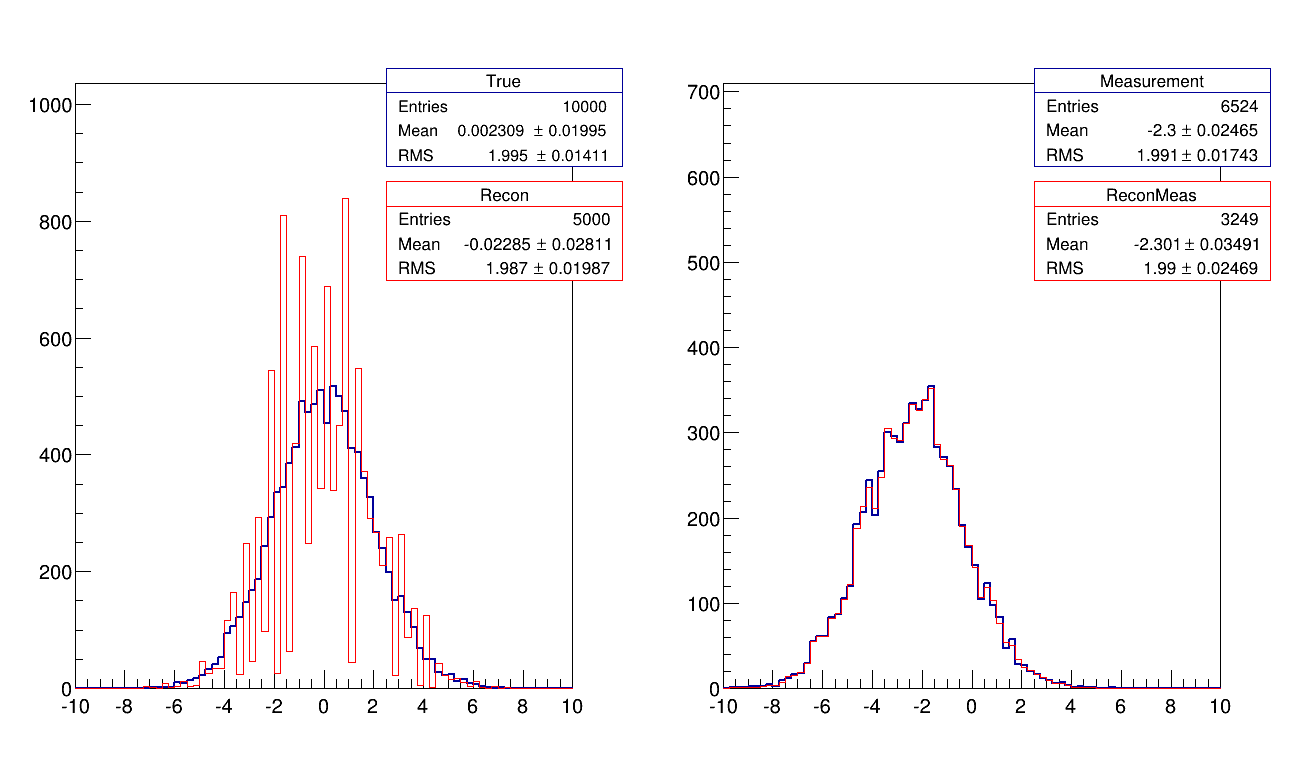}
 \includegraphics[width=.53\textwidth]{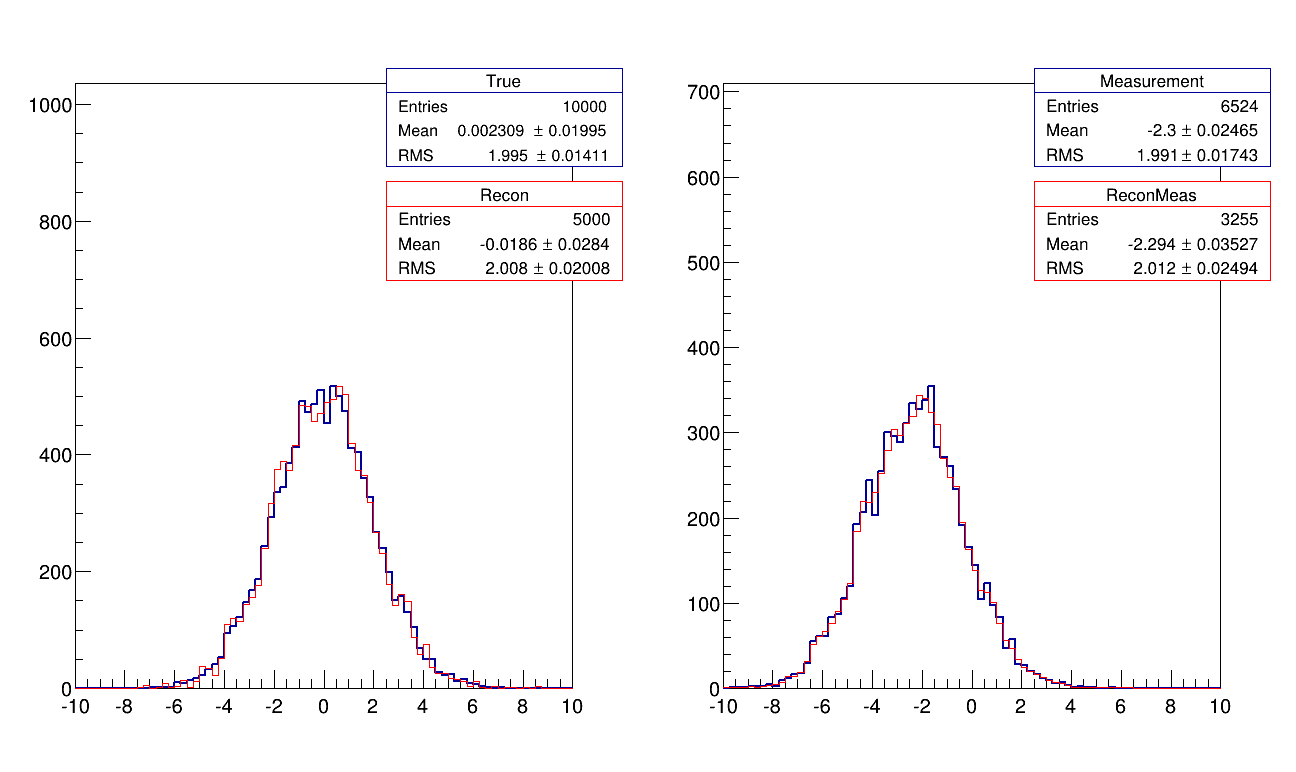}
\end{center}
\caption{Illustration of the contribution of the regularization term in the \textit{selection criteria}.
The progress of the algorithm after 5000 iterations. 
Top: $\chi^2$ test only, no regularization. Bottom: $\chi^2$ test and regularization.}
\label{reg}
\end{figure}

The $\chi^2$ test of the candidates is not sufficient to ensure a good reconstruction of the true distribution.
This problem is well known and comes from the ill-posedness of the matrix $R_{ij}$. As a result, 
the presence of small statistical fluctuations in the measured sample has a very disproportional effect
on the reconstructed distribution. This is illustrated on Fig. \ref{reg} - Top.

The problem can be mitigated by adding a regularization term to the \textit{selection criteria} of the \textit{best guess}:

\begin{equation}
 \min(\chi^2 + \alpha C) ,
\end{equation}
where  $C$ is the regularization term and $\alpha$ is its relative weight in the \textit{selection criteria}.
The role of the regularization term is to add a penalty for guesses $T_j^g$, which give very good matching between
the measured sample $M_i$ and the projection of the guess $M_i^g$, but are nonsensical.
The role of the coefficient $\alpha$ is to ensure that the $\chi^2$ test will dominate the selection
criteria and that the regularization term will add only a weak preference to this criteria.
One possible implementation of the regularization term is:

\begin{equation*}
 {T^g_j}' = 2\frac{\Big( T_j^g/B_j - T_{j-1}^g/B_{j-1} \Big) }{(B_j+B_{j-1})}
\end{equation*}

\begin{equation*}
 {T^g_j}''= {T^g_{j+1}}'-{T^g_j}'
\end{equation*}

\begin{equation} \label{regterm}
 C = \frac{R^4}{\left(\sum_{j=0}^{N_t-1} T^g_j\right)^2} \times \frac{\sum_{j=1}^{N_t-2}({T^g_j}'')^2}{N_t-2}
\end{equation}

Here $T_j^g$ is the number of entries in bin $j$ of the candidate, $B_j$ is the size of the bin $j$
and R is the range of the true sample (difference between the lower edge of the first bin and the upper edge of the last bin).
This regularization term will prefer smooth distributions and will constrain all very complex distributions having large
bin-to-bin fluctuations\footnote{The formulation of this regularization term is a bit complicated, because it tries to handle
the case of non-uniform bin sizes. }. The effect of adding a regularization term in the \textit{selection criteria} is illustrated on
Fig. \ref{reg} - Bottom.

The requirement of having a smooth distribution is not the only possibility for the regularization term. Any additional
information, known in advance, for the true distribution $T$ can be used to define a regularization term. It is also possible to
have multiple regularization terms, having different relative weights in the \textit{selection criteria}. 

\section{2D unfolding}
The implementation of the 2D unfolding requires only a minor modification of the procedure described so far.
The 2D histogram of the true distribution $T_{kl} \ (k=0,1,..,N_t-1; l=0,1,..,M_t-1)$ can be treated as a 1D histogram
$T_{j} \ (j=0,1,..,N_t \times M_t-1)$. The same can be done for the measured distribution $M_{mn}$. The only
considerable difference between 1D and 2D comes from the definition of the regularization term of the \textit{selection criteria}.

\begin{figure}[htb]
\begin{center}
 \includegraphics[width=.53\textwidth]{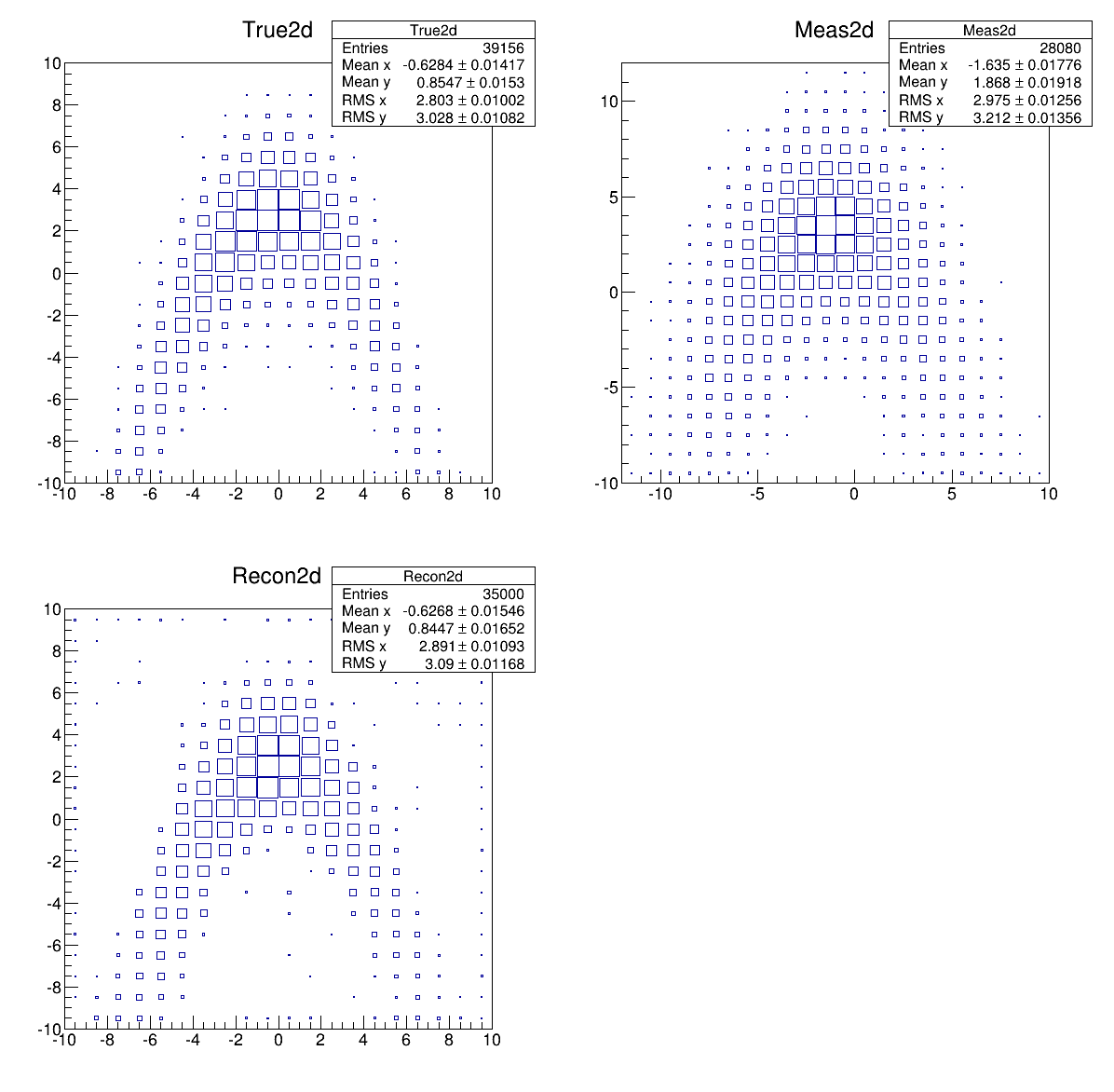}
\end{center}
\caption{Reconstruction of a 2D distribution.
         Top left: the true distribution $T_{kl}$.
         Top right: the  measured distribution $M_{mn}$.
         Bottom left: the reconstructed distribution $T^g_{kl}$
}
\label{2d}
\end{figure}

Fig. \ref{2d} demonstrates the reconstruction of a complex 2D distribution. The regularization term used in this example is similar to 
(\ref{regterm}), but the smoothness of the reconstructed distribution is checked independently in $X$ and $Y$ direction.
As in the 1D example above, here the connection function corresponds to a gaussian smearing, systematic translation, and
variable inefficiency. The elements of $R_{ij}$ are calculated assuming a flat true distribution.

\begin{figure}
\begin{center}
 \includegraphics[width=1.2\textwidth,angle=90]{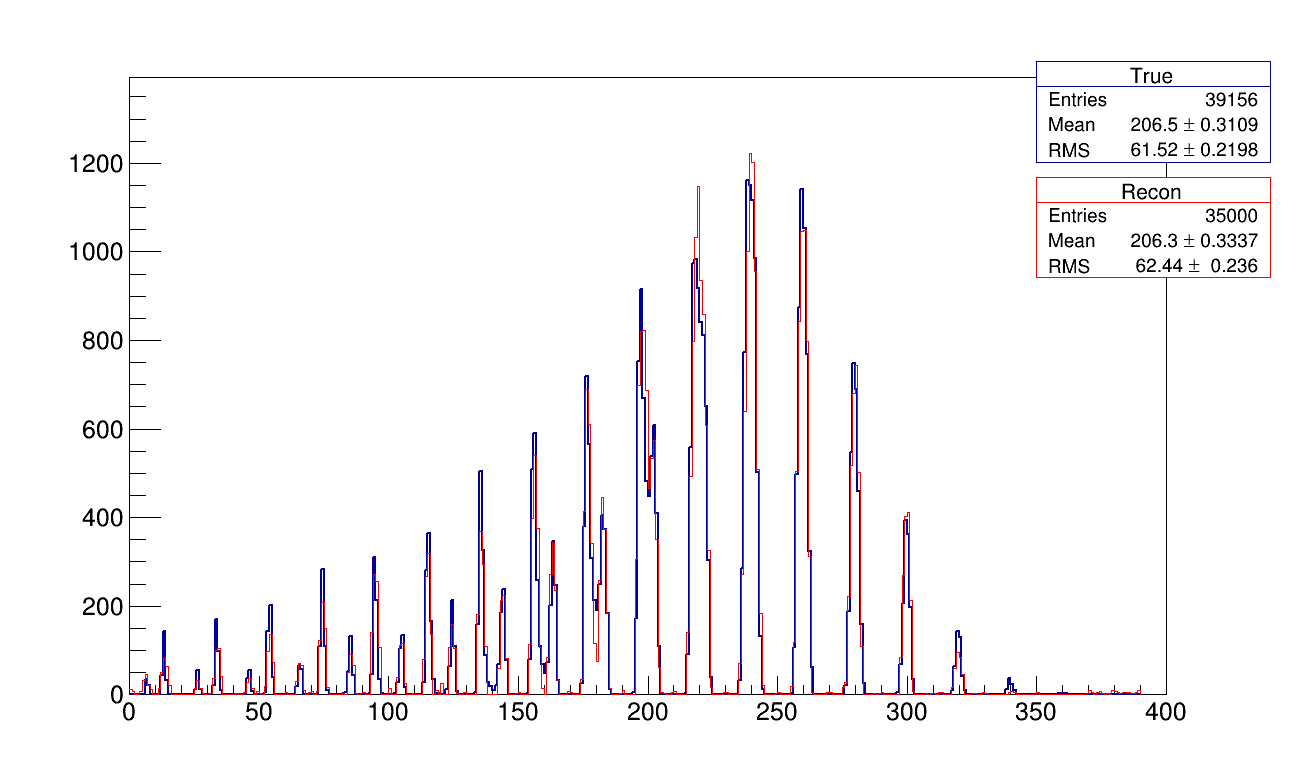}
\end{center}
\caption{Reconstruction of a 2D distribution. The true distribution $T$ in blue and the reconstructed
 distribution $T^g$ (\textit{best guess}) in red, plotted as 1D histograms.}
\label{2d_1d}
\end{figure}

Keep in mind that the algorithm does not reconstruct directly the 2D true distribution $T_{kl}$. What is actually
reconstructed is the 1D true distribution $T_{j}$ (see Fig. \ref{2d_1d}). Notice, that this quite complex 1D distribution is
reconstructed without any initial assumptions\footnote{One may argue that the regularisation term used here is actually an 
initial assumption.}. 

\section{Discussion of the method}
The heuristic method described so far, can be classified as a Greedy\cite{greedy} Genetic\cite{genetic} algorithm.
It does not apply any restrictions on the configuration of the bins used to describe the true and
the measured distributions and on the dimensions of the connection matrix $R_{ij}$. The method itself does not require explicitly
any knowledge about the true distribution, but if we have additional information known in advance, this can be used to define a
regularization term and improve the quality of the solution. 

Nevertheless, the method relies on the good knowledge of the matrix  $R_{ij}$ and any systematic or statistical errors in the
calculation of the matrix elements will affect the quality of the solution.

The realization of the method has been implemented as a small C++ library available at 

https://launchpad.net/ggaunfold

\end{multicols}

\

\end{document}